\documentclass[eqsecnum,aps,showpacs]{revtex4}
\newcommand\beq{\begin{equation}}
\newcommand\eeq{\end{equation}}
\def\beq{\begin{equation}}
\def\eeq{\end{equation}}

\begin{document}

\title{Gravitational Wave Chirp Search:
Economization of PN Matched Filter Bank\\
via Cardinal Interpolation}
\author{R.P. Croce and Th. Demma}
\address{Wavesgroup, D.I.$^{3}$E., University of Salerno, Italy}
\vspace*{-.3cm}
\author{V. Pierro and I.M. Pinto} 
\address{Wavesgroup, University of Sannio at Benevento, Italy}
\vspace*{-.3cm}
\author{D. Churches and B.S. Sathyaprakash}
\address{Department of Physics and Astronomy, Cardiff University, Cardiff, UK}

\begin{abstract}
The final inspiral phase in the evolution of a compact binary consisting
of black holes and/or neutron stars is among the most probable events
that a network of ground-based interferometric gravitational wave   
detectors is likely to observe. Gravitational radiation emitted during
this phase will have to be dug out of noise by matched-filtering 
(correlating) the detector output with a bank of several $10^5$ templates, 
making the computational resources required quite demanding, though not formidable. 
We propose an interpolation method for evaluating the correlation 
between template waveforms and the detector output 
and show that the method is effective in  substantially reducing 
the number of templates required.  
Indeed, the number  of templates needed could be  
a factor $\sim 4$ smaller than required by the usual
approach, when the minimal overlap 
between the template bank and an arbitrary signal 
(the so-called {\it minimal match}) is $0.97$. 
The method is amenable to easy implementation, 
and the various detector projects 
might benefit by adopting it to
reduce the computational costs 
of inspiraling neutron star 
and black hole binary search.
\end{abstract}


\maketitle

\section{Introduction}
Black hole (BH) and neutron star (NS) binaries 
end their lives by emitting a chirp-like burst 
of  gravitational wave (GW) radiation. 
The inspiral radiation  from NS-NS sources 
located within a distance of 20 Mpc, 
and  BH-BH sources within  200 Mpc, should be observable 
by a  first-generation  network of  ground-based  
long-baseline interferometric GW detectors   
\cite{review}.
Correlating  the detector output (data)
with a family of expected waveforms (templates) 
and using the largest correlator as a detection statistic,
provides the best detection strategy  
for Gaussian colored stationary noise 
(maximum likelihood, matched filtering \cite{SigTh}),
but as we will not know the waveform parameters beforehand, 
a bank of 10-100 thousands of templates will have to be used
so as not to miss out any source \cite{Owe_Sat99}.   
Data analysis groups have been, therefore, steadily looking 
for data analysis economization strategies.

In setting up a bank of templates to search for a signal
with uknown parameters 
the quantity of interest \cite{why} 
is the (frequency domain) scalar product
between the signal $h(\cdot)$ and the template(s) $g(\cdot)$. 
The scalar product defined in this context 
is similar to the usual vector space definition, 
except that the measure is derived from the detector
noise power spectral density $S_n(f)$:
\begin{equation}
\langle h, g \rangle \equiv 2 \left [ \int_0^\infty \tilde h(f) \tilde g^*(f)
\frac{df}{S_n(f)} + {\rm C.C.} \right ],
\label{eq:scal_prod}
\end{equation}
where C.C. denotes the complex conjugate.
It is customary to use unit-norm templates,  
for which $||g|| \equiv \left <g, g \right >^{1/2}=1$.
Under this assumption the scalar product is also called
{\em deflection}, and denoted by $d$ \cite{deflec}. 
The normalized deflection obtained  by further dividing
$\left<g, g\right >$ by $||h||$ is the
{\em overlap} or ambiguity function.
It is a real number in $[0,1]$,
and is a measure of how similar the template 
and the signal are.

Given a binary inspiral signal and a template,
the (nominal) times at which they coalesce 
and the phases at the times of coalescence 
are unimportant for the purpose of data analysis.
We can, therefore, preliminarily maximise the overlap 
over these parameters  \cite{how}. 
Such a maximised  overlap is called the {\it match}  
\cite{Owe_96}, and is a function of the remaining
(intrinsic) source and template parameters \cite{NCC}.

The minimal match $\Gamma$ is the smallest {\it overlap} 
of a signal of arbitrary parameters 
with the template closest to it in the template bank \cite{FF}.  
It is immediately related to the fraction $(1-\Gamma^3)$
of  potentially observable sources  
which {\em might}  be  missed out \cite{Owe_Sat99}.

In a recent paper \cite{verbum} it has been pointed out
that (under certain assumptions) the match is a quasi
band-limited function of the difference between the
intrinsic source and template parameters.
As such, it can be {\em approximated} by a cardinal
imterpolating expansion, which uses only
{\em a fraction} of the templates needed by the
std. lattice for a prescribed minimal match.   
The same {\em economized} template set can
be used to build a cardinal-interpolated formula
for the partially maximised correlator, under the
same minimal match constraint.

This was shown in \cite{verbum} 
for the simplest (but rather unrealistic) case 
of Newtonian (0PN) chirp signals. 
At the lowest Newtonian (0PN) order 
an inspiral waveform depends only on a single parameter \cite{chirpmass}.
In this one-parameter template space, 
the cardinal-interpolation method \cite{verbum} leads to
a reduction in the number of templates by a factor $\approx 1.4$
for $\Gamma=0.97$, and
the resulting template density is close 
to an absolute minimum 
set by the theory of quasi-band-limited functions.  
As anticipated in  \cite {verbum}, 
one expects an even {\em larger} reduction 
in the number of templates 
for multi-parameter PN models, 
in view of well known properties 
of cardinal interpolating expansion in several dimensions 
(see references quoted in \cite{verbum}).

In this Rapid Communication we present results 
for the case of  first-post-Newtonian (1PN) signals and templates, 
which support the above expectation. 
We show that in the two-dimensional  1PN template space  
the reduction in the number of templates is by a factor $\approx 4$
for $\Gamma=0.97$.
A similar reduction in the computational resources is implied,
which we believe the interferometric detector 
projects such as TAMA, GEO, LIGO and VIRGO might benefit from.  
In the following we present only essential ideas and results, 
deferring the details to a longer version. 
Throughout this paper we use geometrized units (i.e., $G=c=1$).

\section{The post-Newtonian Waveform}
We shall use the {\em restricted} PN approximation,
where the amplitude of the waveform is kept to lowest
(Newtonian) PN order, while
the phase is expanded to the highest PN
order available \cite{restricted}.

The evolution of the instantaneous orbital phase $\varphi$ 
of an (adiabatically) inspiraling compact circular-orbit binary
is obtained (in parametric form)
by solving a pair of ordinary differential equations \cite{Dam_etal98}:
\begin{equation}
\frac{dt}{dv} = -\frac{m}{F} \frac{dE}{dv},\ \ \ \   
\frac{d\varphi}{dv} = - \frac{v^3}{F} \frac{dE}{dv},    
\label{eq:ODE}
\end{equation}
where $v$ is the gauge independent relative velocity of the two stars,
${F}(v)$ is the gravitational wave flux (luminosity),
and $E(v)$  the dimensionless relativistic binding energy of the system,
related by $F(v)=-m(dE/dt)$.
The  instantaneous gravitational wave frequency $f_{GW}$  
is related to the orbital phase, 
and hence via (\ref{eq:ODE}) to $v$, by:
\begin{equation}
f_{GW}=\pi^{-1} \frac{d\varphi}{dt} = \frac{v^3}{\pi m} .
\label{eq:fGW_vs_v}
\end{equation}
The flux and energy functions are presently known
to order $v^5$ (2.5 PN) for non-spinning binaries \cite{Blanchet}.

The gravitational wave
(dominant, $m=2$ multipole, $2nd$ harmonic of orbital frequency)  
emitted by an inspiraling compact binary and sensed
by an interferometric antenna is described by the waveform \cite{signal}:
\begin{equation}
h(t) = 4C\eta 
\frac {m}{r}  
v^2(t) \cos[2\varphi(t)],   
\label{eq:TD_signal}
\end{equation}
where $r$ is the distance to the source, 
and $C$ is a constant in the range $[0,1]$ 
dependent on the relative source/detector orientation 
with r.m.s  value $2/5$ 
(average over all orientations and wave polarisations) \cite{directivity}.

It is straightforward to obtain an explicit frequency-domain 
(Fourier transform) representation of the waveform ~(\ref{eq:TD_signal})
using the stationary phase approximation 
\cite{Sat_Dhu91,SP_discuss}:
\begin{equation}
\tilde h(f) =
2C\eta 
\frac{m}{r}  
\frac{v^2(t_f)}{\sqrt{\dot{f}_{GW}(t_f)}}   
e^{i [\psi_f(t_f) - \pi/4]},
\label{eq:FD_signal}
\end{equation}
where $t_f$ is the stationary point \cite{stat_phas}
and an overdot denotes the derivative w.r.t. $t$.
On substituting for the stationary point $t_f$, 
and consistently using the available PN expansions 
of the flux and energy functions, 
the phase  in Eq.~(\ref{eq:FD_signal}) 
takes on the simple form:
\begin{equation}
\psi_f(t_f) = 2\pi f T_C - \phi_C +2
\int_{v_f}^\infty dv
(v_f^3-v^3)\frac{dE/dv}{F}=
2\pi f T_C - \phi_C + \sum_0^{4} \Psi_k(f) \tau^k,
\label{eq:phasing}
\end{equation}
where $v_f \equiv v(t_f) = (\pi m f)^{1/3}$,
$T_C$ is the (nominal) 
time of coalescence \cite{Tc_warn}, 
$\phi_C$ is the phase at $t=T_C$.
The $\tau^k$ are the so-called (dimensionless)  
PN   {\it chirp times}  \cite{Sat_94,PN_note}:
\begin{equation}
\tau^0 = \frac{5}{256\eta v_0^5}, \  
\tau^1=0, \ 
\tau^2=\frac{5}{192\eta v_0^3}   
\left ( \frac{743}{336} + \frac{11}{4} \eta\right ),  \ 
\tau^3 = \frac{\pi}{8\eta v_0^2}, \ 
\tau^4=\frac{5}{128\eta v_0}   
\left ( \frac{3058673}{1016064} + \frac{5429}{1008} \eta +   
\frac{617}{144}\eta^2 \right),
\end{equation}   
where $v_0=(\pi m f_0)^{1/3}$,
$f_0$ is a scaling frequency  (to be specified below), 
and the $\Psi_k$ are given, 
in terms of the scaled frequency  $\nu \equiv f/f_0$, by:
\begin{equation}   
\Psi_0 = \frac{6}{5\nu ^{5/3}},\   
\Psi_1=0,\ 
\Psi_2 = \frac{2}{\nu },\   
\Psi_3 = \frac{-3}{\nu ^{2/3}},\   
\Psi_4 = \frac{6}{\nu ^{1/3}}.
\end{equation}   
In the  {\em restricted}  PN approximation 
the amplitude  in (\ref{eq:FD_signal}) 
depends on frequency simply through a factor $f^{-7/6}$.

\section{The First post-Newtonian Match and the Standard Template Bank}
In the stationary phase restricted PN approximation for $\tilde h$,
the match (or {\it reduced normalized deflection}, 
in the jargon of \cite{verbum}), can be written:
\begin{equation}
\bar{D} = \max_{\Delta T_{c}}   
\frac{
\displaystyle{
\left| 
\int_{f_{\rm i}}^{f_{\rm s}}
df~\frac{f^{-7/3}}{S_n(f)}
e^{i(2\pi f\Delta T_{c}+\Delta \Psi )} 
\right| 
}
}{
\displaystyle{
\int_{f_{\rm i}}^{f_{\rm s}}
df~\frac{f^{-7/3}}{S_n(f)}
}},
\label{eq:dbar}
\end{equation}
where $[f_{\rm i},f_{\rm s}]$ is the antenna spectral window, 
$\Delta T_{c}$ is the difference in the coalescence times 
of the signal and template,   
and, at the 1PN level of approximation,
\begin{equation}
\Delta \Psi(\nu) = \frac{6}{5\nu ^{5/3}}\Delta \tau^0 +
\frac{1}{\nu } \Delta \tau^2,    
\label{eq:deltapsi}
\end{equation}
where $\Delta \tau^{i}\!=\!\tau^{i}_{S}\!-\!\tau ^{i}_{T}$, $i=0,2,$ 
are the differences in the chirp times of the source ($S$) 
and  template ($T$), respectively,
$\nu=f/f_0$, and $f_0$ is the frequency at which $S_n(f)$ is minimum.
In the sequel we shall adopt the LIGO form \cite{LIGO_noise}
of the (one-sided) noise power spectral density,   
\begin{equation}
S_n(\nu)=\frac{S_{0}}{5}\left[ \nu^{4} + 2\left( 1+\nu^2 \right) 
\right].    
\label{eq:PSD}
\end{equation}
The reduced deflection Eq.~(\ref{eq:dbar}) depends on $f_{\rm i}$, $f_{\rm s}$   
and $f_{0}$ only through the integration limits, which become 
$\nu_{\rm i}\equiv f_{\rm i}/f_{0}$ and $\nu_{\rm s}\equiv f_{\rm s}/f_{0}$.

Following Owen \cite{Owe_96} 
we shall perform a coordinate transformation which,
asymptotically in a neighbourhood of the peak $\Delta\tau^0=\Delta\tau^2=0$,
re-shapes the surface $z=\bar{D}$ into a convex circular paraboloid:   
\begin{equation}
\left\{   
\begin{array}{l}
\Delta{\tau}^0=\Delta x^0 \cos\vartheta - \Lambda\Delta x^1
\sin\vartheta, \\   
~ \\   
\Delta{\tau}^2=\Delta x^0 \sin\vartheta + \Lambda\Delta x^1
\cos\vartheta.
\end{array}
\right.  \label{eq:trasfo}
\end{equation}
The above transformation depends on the choice 
of $f_0$, $\nu_{\rm i}$ and $\nu_{\rm s}$. 
In the following, for illustrative purposes, 
we let $f_0=200$~Hz  (initial LIGO), 
$\nu_{\rm i}=0.2$ and $\nu_{\rm s}=4$, 
whereby $\vartheta \approx 0.489$ and $\Lambda=15.203$. 
The resulting function $\bar{D}(\Delta x^0, \Delta x^1)$
is shown in Fig.~1.

In the standard lattice approach, the template spacing $\delta_L$ 
is obtained by enforcing the minimal match condition, 
whereby for {\em any} admissible source, 
there exists {\em at least one} template such that $\bar D \geq \Gamma$. 
The curves $\bar{D}=\Gamma$, shown in the inset of Fig.~1,
are {\em nearly} circular \cite{foot1}
down to $\Gamma = 0.95$, which includes all values 
of practical interest for a single-step search.
Thus, $\delta_L$ is the side-length of the inscribed square. 
In the inset of Fig.~2 we display $\delta_L$ as a function of $\Gamma$.

\section{The Cardinal Interpolation for the 1-PN  Match}
In the cardinal-interpolation approach  the
template spacing $\delta_C$ is obtained by enforcing
the minimal match condition on the cardinal expansion of the match \cite{verbum}.
The $2D$ cardinal  expansion of the 1PN match is: 
\begin{equation}
\bar{D}_B(\Delta x^0, \Delta x^1) = \sum_{m,n}^{-\infty,+\infty} 
\bar{D}_B(\Delta x^0_m, \Delta x^1_n)~
\mbox{sinc} \left[ \frac{\pi}{\delta_C}\left(\Delta x^0-\Delta x^0_m\right) \right]
\mbox{sinc} \left[\frac{\pi}{\delta_C} \left( \Delta x^1-\Delta x^1_n \right) \right],
\label{eq:cardex}
\end{equation}
where $\Delta x^0_{m+1} \! - \! \Delta x^0_m\! = \! \Delta x^1_{n+1}\! -\! \Delta x^1_{n}=
\delta_C$ is the cardinal spacing to be determined.
This expansion represents {\em exactly} \cite{verbum} the function:   
\begin{equation}
\bar{D}_B = {\cal F}^{-1}_{[\Delta \vec{y} \rightarrow \Delta \vec{x}]}
\left\{ W \left( \frac{\Delta\vec{y}}{B}\right)~ {\cal F}_{[\Delta \vec{x}
\rightarrow \Delta \vec{y}]} \bar{D}(\Delta\vec{x}) \right\},
\end{equation}
where ${\cal F}$ is the Fourier-transform operator,    
\begin{equation}
W \left( \frac{\Delta y^0}{B},\frac{\Delta y^1}{B} \right)= \left\{   
\begin{array}{l}
1,~~~|\Delta y^0| < B, |\Delta y^1| <B, \\   
~ \\   
0,~~~\mbox{elsewhere,}
\end{array}
\right.
\end{equation}
and $B=(2\delta_C)^{-1}.$
Similar to the Newtonian case \cite{verbum}, 
the 1PN match Eq.~(\ref{eq:dbar}) 
is a quasi-band-limited function in the $L^\infty$ norm, namely,   
\begin{equation}
\exists \gamma, B_c \in {\cal R}^+:\sup |\bar{D}-\bar{D}_B| = \exp[
-\gamma(B-B_c)].  
\label{eq:experr}
\end{equation}
The cardinal expansion Eq. (\ref{eq:cardex}) 
can be shown to {\em approximate}
the function Eq. (\ref{eq:dbar}) 
in the $L^{\infty}$ (and $L^{2}$) norm, 
using the {\em minimum} 2D sample density $1/\delta_C^2$ 
compatible with a prescribed accuracy 
(see \cite{verbum}, and references cited therein). 
The {\it exponential} decay of the error 
(\ref{eq:experr}) is exemplified in Fig.~3.

We shall now discuss the template spacing to be used   
in Eq. (\ref{eq:cardex}), for a prescribed minimal match $\Gamma$.

The source and template parameters can be always conveniently written as
follows:   
\begin{equation}
\vec{x}_{S}=\vec{x}_{Q}+\vec{\alpha}_{S}\delta_C, \ \   
\vec{x}_{T}=\vec{x}_{Q}+\vec{\alpha}_{T}\delta_C,
\label{eq:alphas}
\end{equation}
where $\alpha_{S,T}^{0,1} \in [0,1]$ and
$\vec{x}_{Q}$ correspond to the lower left corner of a suitable
template-space cell.

For any given value of $\delta_C$, as the source coordinates $\vec{x}_{S}$
move in the template-space cell, the location $\vec{x}_{T}^{\rm max}$ where
the cardinal-interpolated match $\bar{D}_B$ attains its maximum value   
and the maximum value itself $\bar{D}_{B}^{\rm max}$ change.
The minimal match condition should be enforced 
for the special value $\vec{\alpha}_{*}$ of   
$\vec{\alpha}_{S}$ where the interpolated {\em maximum} is a {\em minimum}.
   
Numerical experiments show that
$\vec{\alpha}_{*}=(0.5,0.5)$, 
which corresponds to the center of the cell,
below some critical value $\delta_\times$ of $\delta_C$. 
Above this critical value, a bifurcation occurs, i.e.,
{\em two} (equal) minima exist \cite{foot2} 
located on the descending diagonal of the cell, 
where   
\begin{equation}
\alpha^0_{S}=0.5+\xi,~~~\alpha^1_{S}=0.5-\xi,  \label{eq:pos}
\end{equation}
symmetrically with respect to the cell center, viz.   
at:
\begin{equation}
\xi = \pm \xi_{*}(\delta_C),~~~~~\xi_{*}(\delta_C) \in [0,0.5].  \label{eq:diag}
\end{equation}
This is illustrated in Fig.~4, 
where we show the density plots of   
$\bar{D}_B^{\rm max}$ as a function of $\vec{\alpha}_{S}$ 
in the fundamental cell, for the representative cases 
$\delta_C =0.078 < \delta_\times$ and   
$\delta_C = 0.099 > \delta_\times$.

The values of $\bar{D}_B^{\rm max}$ on the descending diagonal of the cell,
where the minima are located and Eq.~(\ref{eq:pos}) holds, are plotted in Fig.~5,
as a function of the displacement $\xi$ off the cell center, for several values
of $\delta_C$. The values of $\bar{D}_B^{\rm max}$ at the minima of these curves,
occurring at $\xi=\xi_{*}(\delta_C)$, are the minimal matches.

It is seen, e.g., that for $\Gamma=0.97$ one has $\delta_C \approx 0.104$.
By comparison with Fig.~2, where $\delta_L=0.052$ for $\Gamma=0.97$, we
deduce that $\delta_C/\delta_L \approx 2$. Hence the reduction in the 2D
template density is of the order of $(\delta_C/\delta_L)^2 \sim 4$, which is
quite remarkable.
It is also seen that, similar to the 0PN case, the 1PN template density
reduction with respect to the plain lattice increases with   
$\Gamma$, as shown in Fig.~2.

\section{Conclusions}
The number of 1PN templates required to keep the minimal match above a
given threshold $\Gamma$ 
can be significantly (by a factor $\approx 4$ at $\Gamma \approx 0.97$) 
reduced by using cardinal interpolation. 
The statistical properties of the 1PN
cardinal-interpolated (partially maximised) correlator bank   
will be the subject of a forthcoming paper.   
Extension to  2.5PN templates should be straightforward, in principle,
using the (almost-flat parameter space manifold)
coordinates introduced by Tagoshi and Tanaka  
\cite{TaTa}.

\section*{Acknowledgements}

This work has been sponsored in part by the EC through a senior
visiting scientist grant to I.M. Pinto at NAO, Tokyo, JP.
I.M. Pinto wishes to thank the TAMA staff at NAO   
and in particular prof. Fujimoto Masa-Katsu and prof. Kawamura Seiji
for gracious hospitality and stimulating discussions. D. Churches 
and B.S. Sathyaprakash  thank R. Balasubramanian for helpful discussions.


\newpage
\Large
\begin{center}
{\bf Figure Captions}
\end{center}
\normalsize
$$~$$
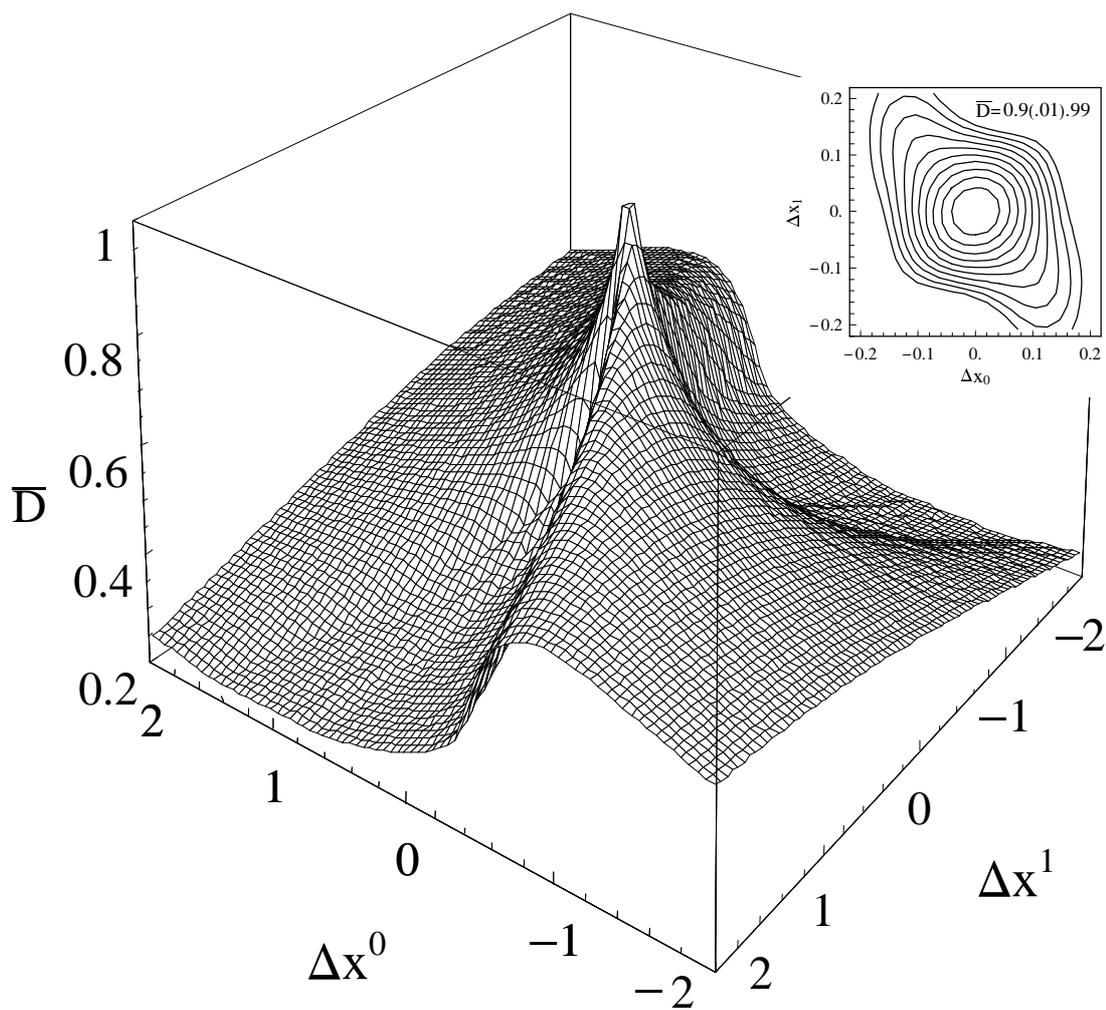
Fig. 1 - The function $z=\bar{D}(\Delta x^0,\Delta x^2)$, and some
of its contour levels (inset).
$$~$$
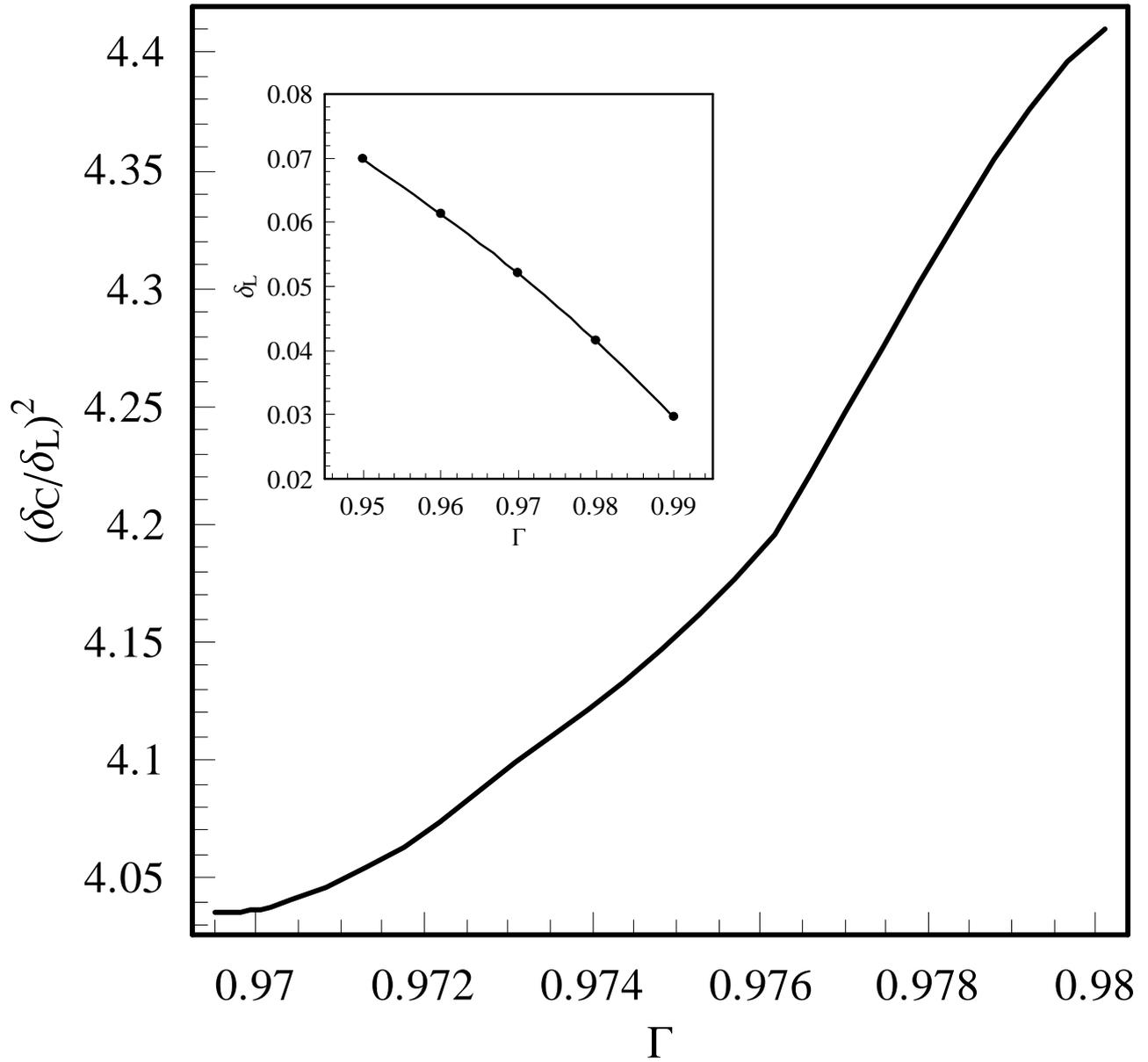
Fig. 2  - The plain 1PN lattice spacing $\delta_L$   
(inset), and the $2D$ template density reduction $(\delta_C/\delta_S)^2$,   
achieved by use of cardinal interpolation, 
both plotted as functions of the match $\Gamma$.
$$~$$
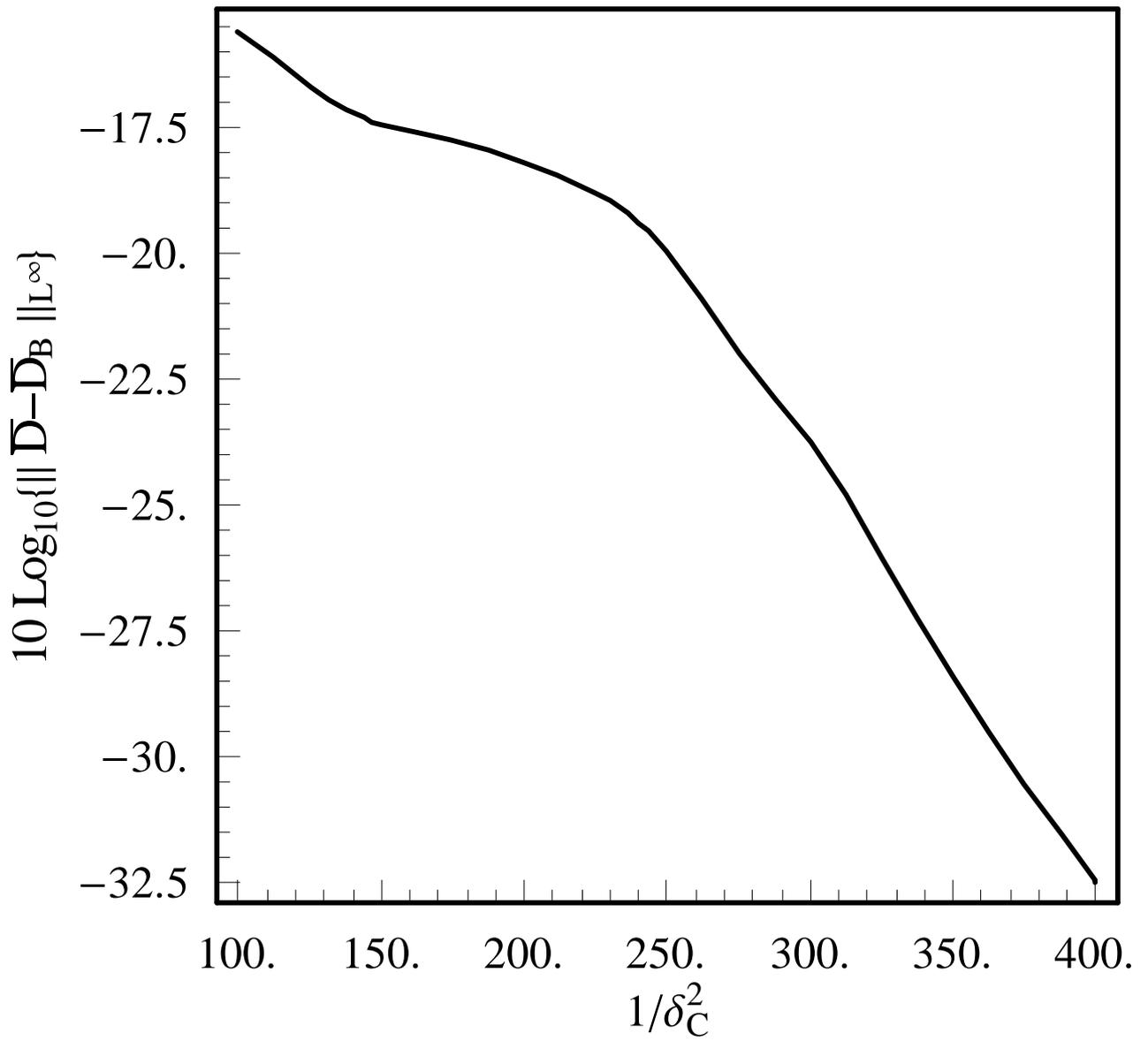
Fig. 3 - The L$^\infty$ error  as a function of $\delta_C^{-2}.$
$$~$$
Fig. 4 - Density plot of $\bar{D}^{\rm max}$ vs. $\vec{\alpha}_S$ for   
$\delta_C=0.0078$ (left) and $\delta_C=0.0099$ (right).
$$~$$
Fig. 5 - $\bar{D}_B^{\rm max}$ vs. $\xi$, for several values of $\delta_C$.
$$~$$

\begin{thebibliography}{99}
\bibitem{why}{Under the given assumptions,
the correlators obtained from the scalar
product between the noisy data and the templates  
will be random gaussian, with mean (\ref{eq:scal_prod}).}
\bibitem{deflec}{By Schwartz inequality, $d \leq ||h||^{1/2}$, 
which is the signal-to-noise ratio (SNR), 
the equality holding iff $g \propto h$.}
\bibitem{how}{The maximisation is readily accomplished by taking the maximum
absolute value of the (inverse) Fourier transform of  $\tilde h(f) \tilde g(f) /S_n(f)$ \cite{Owe_96}.}
\bibitem{NCC}{The partially maximised correlator
(scalar product between the noisy data and the template(s)) is
called reduced ($\Delta T_C$ maximised) non coherent
($\Delta\phi_C$ maximised) correlator in \cite{verbum}.}
\bibitem{chirpmass}{The referred parameter is ${\cal M}^{-5/3}$,
where ${\cal M}=\eta^{3/5} m$ is the chirp-mass, 
$\eta=m_1m_2/m^2$ being the symmetric mass ratio, 
$m\!=\!m_1\!+\!m_2$ the total mass of the binary, and
$m_1$ and $m_2$ the masses of the component stars.}
\bibitem{stat_phas}{Here the relevant phase 
is  $\psi_f(t) \! = \! 2\pi f t \! - \! 2\varphi(t)$, 
and the stationary point $t_f$ is defined by 
$ {2\pi}f \! = \! 2\dot{\varphi}(t_f) \! = \! 2\pi f_{GW}(t_f)$.}
\bibitem{PN_note}{Note that $\tau^1=\Psi_1=0$, because of the absence of
$O(v/c)$ terms in the PN expansion.}
\bibitem {foot2} {This peculiar behaviour is due to the already mentioned fact 
that under the transformation Eq.~(\ref{eq:trasfo}) the curve $\bar{D}=\Gamma$ is {\em not exactly} a circle. Indeed, if this were the case, the minimum would always
be at $\vec{\alpha}_{S}=\vec{\alpha}_{*}=(0.5,0.5)$.}
\end{thebibliography}
\end{document}